# Nonlocal generalization of the Chapman-Enskog method and hydrodynamic equations. Nonlocal vorticity transport in a turbulent medium

A. B. Kukushkin[a,b,*]

[a]*National Research Center "Kurchatov Institute," Moscow, Russia*
[b]*National Research Nuclear University MEPhI, Moscow, Russia*
**e-mail: Kukushkin_AB@nrcki.ru; kukushkin.alexander@gmail.com*

**Аннотация**
The results from [Fluid Dyn 61, 20 (2026)] — concerning the generalized Chapman-Enskog method and a nonlocal (i.e., non-diffusive, spatially integral) viscous stress tensor in the Navier-Stokes equation — are extended to the case of arbitrary gradients of not only hydrodynamic velocity but also temperature. The resulting equations simplify the description of tangential discontinuities and separated flows. For the specific case of an incompressible medium, a nonlocal vorticity transport equation is derived that is structurally identical to the Biberman-Holstein equation for resonant radiation transport in plasmas and gases. A power-law form has been determined for the velocity distribution function of long-free-path vorticity carriers that satisfies empirical Richardson's $t^3$ law for the mean square of relative separation of a pair of particles in a turbulent liquid or gaseous medium.

## 1. Introduction

The Chapman-Enskog method [1] was generalized in [2] to the case of significant hydrodynamic velocity gradients. This made it possible to derive a representation of the viscous stress tensor in the Navier-Stokes equation that is integral with respect to spatial coordinates. In the limit of short free paths of medium perturbations, the tensor reduces to the standard form, which, as is well known, is difficult to apply to the description of tangential discontinuities and separated flows. In the present work, this result is generalized to the case of arbitrary gradients of not only the hydrodynamic velocity but also the temperature (Section 2).

The next step was to derive an equation for the vorticity (the curl of the hydrodynamic velocity), as this is of particular interest to turbulence theory given the well-known role of vortex structures — a significance dating back to Kolmogorov's theory [3] of homogeneous isotropic turbulence in an incompressible fluid. Within this framework, an equation for the nonlocal (non-diffusive) transport of vorticity is derived from equation [2] (Section 3); this equation proved to be structurally identical to the Biberman-Holstein equation [4–7] describing resonant radiation transport in plasmas and gases.

This step made it possible to determine the form of the velocity distribution function for long-free-path vorticity carriers that satisfies empirical Richardson's $t^3$ law [8] for the mean square of relative separation of a pair of particles in a turbulent liquid or gaseous medium (Section 4). It is worth recalling that the attempt [9] to propose a formalism based on a linear nonlocal (superdiffusive) equation for the adequate description of hydrodynamic turbulence (and, consequently, to ensure that the equation's Green's function matches Richardson's law) was one of the important incentives for the work [2].

## 2. Generalization of [2] to the case of arbitrary temperature gradients

We begin by generalizing equation (2.5) from [2]. Accounting for the temperature gradient (cf. (10.7) in [10]) yields the following equation for the correction $\varphi(\boldsymbol{p},\boldsymbol{r},t)$ — defined in (2.4) of [2] — to the thermodynamically equilibrium distribution function $f^{(0)}(\boldsymbol{p},\boldsymbol{r},t)$ in phase space:

$$f^{(0)}\left[\frac{\partial v_{0i}}{\partial r_k}\left(V_iV_k-\frac{1}{3}\delta_{ik}V^2\right)\frac{m}{T}+\left(\frac{mV^2}{2T}-\frac{5}{2}\right)\frac{\partial lnT}{\partial r_k}V_k\right] + f^{(0)}\left(\frac{\partial\varphi}{\partial t}+v_k\frac{\partial\varphi}{\partial r_k}+F_k\frac{\partial\varphi}{\partial p_k}\right)=-f^{(0)}\frac{\varphi}{\tau}, \quad (1)$$

where $\boldsymbol{V}=\boldsymbol{v}-\boldsymbol{v_0}(\boldsymbol{r},t)$ is the velocity of propagation of the medium disturbance relative to the mean mass velocity of the medium; the remaining notation is given in [2]. Then, for the viscous stress tensor, we ultimately obtain an additional term within the expression in square brackets in (2.13) of [2]:

$$\sigma_{ik}=\frac{4nT}{v_T}\int_0^\infty\frac{x^5dx}{\pi^{3/2}}e^{-x^2}\int\frac{d\boldsymbol{r}'}{|\boldsymbol{r}-\boldsymbol{r}'|^2}n_in_k\left[\frac{\partial v_{0j}}{\partial r_l}\left(n_jn_l-\frac{\delta_{jl}}{3}\right) + \frac{\left(x^2-\frac{5}{2}\right)}{2x}\frac{\partial lnT}{\partial r_j}n_jv_T\right]exp\left(-\int_{r'}^{r}\frac{(\boldsymbol{n},d\boldsymbol{r}'')}{l(x,\boldsymbol{r}'')}\right), \quad (2)$$

where $\boldsymbol{x^2}=\frac{\boldsymbol{mV^2}}{\boldsymbol{2T}}$, $v_T=\sqrt{\frac{\boldsymbol{2T}}{\boldsymbol{m}}}$ is the thermal velocity, integration is carried out over the entire volume of the medium, and the unit vector $\boldsymbol{n}$ corresponds to the solid angle of integration over the coordinate:

$$\boldsymbol{n}=\frac{\boldsymbol{r}-\boldsymbol{r}'}{|\boldsymbol{r}-\boldsymbol{r}'|}.$$

This term represents a thermoforce, the existence of which is well known, for example, in Braginskii's two-fluid hydrodynamics [11] within plasma physics.

Accounting for nonlocality also leads to a generalization of the temperature equation, which has the form [10]:

$$\frac{3}{2}n\left(\frac{\partial T}{\partial t}+v_{ok}\frac{\partial T}{\partial r_k}\right)=-div(\boldsymbol{q})-\frac{\partial v_{0i}}{\partial r_k}(\delta_{ik}nT-\sigma_{ik}), \quad (3)$$

In it, in addition to the change in the viscous stress tensor (cf. (2.12) in [2]), the heat flux density vector,

$$q_i(\boldsymbol{r},t)=\int\varphi(\boldsymbol{p},\boldsymbol{r},t)f^{(0)}(\boldsymbol{p},\boldsymbol{r},t)V_i\frac{1}{2}mV^2d\boldsymbol{p}, \quad (4)$$

must also be modified. This yields the expression for $q_i$ given in (2), in which the tensor preceding the square brackets must be replaced by a vector:

$$\underbrace{n_in_k}\rightarrow n_iv_T\frac{x}{2}. \quad (5)$$

In the limit of short free paths, the resulting expression for the heat flux density reduces to an expression that — up to a numerical coefficient of order unity — coincides with the well-known result of the Chapman-Enskog method:

$$q_i(\boldsymbol{r},t) = \frac{5}{2}\frac{nT\tau}{m}\frac{\partial T}{\partial r_i}. \tag{6}$$

**3. Nonlocal vorticity transport equation**

Equation (2.13) for the viscous stress tensor, derived in [2], makes it possible to obtain a nonlocal vorticity transport equation for the simplest case — namely, a homogeneous isotropic incompressible medium — for which many theoretical results have been obtained, starting with Kolmogorov's theory [3].

In deriving the equation for vorticity — standardly defined by the relation

$$\omega_i(\boldsymbol{r},t) = \varepsilon_{ijk}\frac{\partial v_{0k}}{\partial r_j},. \tag{7}$$

where $\varepsilon_{ijk}$ is the antisymmetric Levi-Civita tensor, account was taken of the fact that the contribution from boundary conditions arising during integration by parts can be described as a source term $B_i$ in the transport equation. Ultimately, the transport equation differs from the standard form in that the vorticity diffusion operator is replaced by an operator representing the balance of generation and destruction of the vorticity $\omega_i(\boldsymbol{r},t)$; in the limit of short free paths, this operator reduces to the aforementioned standard diffusion operator:

$$\frac{\partial \omega_i}{\partial t} + v_{0k}\frac{\partial \omega_i}{\partial r_k} = \omega_k\frac{\partial v_{0i}}{\partial r_k} + \left\{\frac{1}{\tau}\int d\boldsymbol{r}'\,\omega_i(\boldsymbol{r}',t)G(|\boldsymbol{r}-\boldsymbol{r}'|) - \frac{\omega_i}{\tau}\right\} + B_i, \tag{8}$$

where integration with respect to the coordinates is performed over the entire volume of the medium, and the kernel of the integral operator has the form:

$$G(R) = \int_0^\infty \frac{2x^4 dx}{\pi^{\frac{3}{2}}} e^{-x^2}\frac{1}{R^3}\left(e^{-\frac{R}{l(x)}} - 3\left(\frac{R}{l(x)}\right)^3\int_{\frac{R}{l(x)}}^\infty e^{-z}\frac{dz}{z^4}\right), \qquad l(x) = xv_T\tau. \tag{9}$$

The integral of the kernel over the entire space equals unity; this leads to the cancellation of terms when the vorticity is independent of coordinates, whereas accounting for vorticity inhomogeneity results in a diffusion-type differential operator. In the limit of short free paths ($l \to 0$), the vorticity diffusion coefficient coincides with $\eta/\rho = nT\tau/\rho$, a value obtained via our method in the equation for the average mass velocity (see (2.14) in [2]).

Thus, in the nonlocal vorticity transport equation (8), the operator in the braces is structurally identical to the corresponding operator representing the balance between the gain of medium excitation (due to photon absorption by unexcited atoms or ions) and the loss of medium excitation (due to photon emission by excited atoms or ions) in the Biberman-Holstein equation [4–7] for resonant radiation transport in plasmas and gases. This correspondence suggests that the operator $G(|\boldsymbol{r}-\boldsymbol{r}'|)$ describes a similar transport process, but one involving the transport of medium excitation not by photons, but by hydrodynamic vortex structures; consistent with the conditions under which the results in [2] and the present work were obtained, these structures possess a long free path (discussed in greater detail in the next section) and a carrier velocity that is high compared to the mean mass velocity of the medium.

## 4. Conditions for the fulfillment of empirical Richardson $t^3$ law by vorticity carriers in a turbulent medium

Equation (8) makes it possible to determine the form of the velocity distribution function for long-free-path vorticity carriers under which the empirical Richardson $t^3$ law [8] — describing the mean square of separation of a pair of particles in a turbulent liquid or gaseous medium — holds true. To do this, one must refer to the comments in the final paragraph of the text [2]. Indeed, the nonlocal expression for $\sigma_{ik}$ — in which the mean free path characterizes particle pair collisions (such as those between gas molecules) — differs from the standard differential operator only within a narrow layer (i.e., the region of the tangential discontinuity). The Holstein function *T(R)* in the theory of resonant radiation transport, defined as

$$G(R) = \frac{-1}{4\pi R^2}\frac{dT(R)}{dR}, \tag{10}$$

which describes the probability that a carrier of a medium disturbance travels a distance of at least R without being absorbed (i.e., T(R) describes the distribution of the disturbance carriers over free path length), turns out to be an exponential function in the case of disturbance transport by atomic- and molecular-scale particles of the medium,

$$T(R) \propto \exp\left(-3\left(\frac{R}{2v_T\tau}\right)^{\frac{2}{3}}\right), \qquad \frac{R}{v_T\tau} \gg 1. \tag{11}$$

For nonlocal transport, the Holstein function must exhibit a slow, power-law decay at distances far exceeding the characteristic free path (which, in the theory of resonant radiation transport, corresponds to the free path at the spectral line center; see [6, 7]). Therefore, the transport of vorticity or other medium perturbations under discussion is not nonlocal but purely diffusive, although a nonlocal viscous stress operator must be employed in regions where the mean mass velocity is discontinuous. However, if the kinetic equation (2.5) in [2] is treated as an equation describing the transport of macroscopic medium perturbations, and one seeks the thermodynamically equilibrium velocity distribution function for these carriers, then calculating the Holstein function makes it possible to determine the velocity dependence of this distribution function. In this case, the integrals with respect to the variable *x* in (9) will involve for large *x* a power-law function

$$f^{(0)}(\boldsymbol{p}) \propto \frac{1}{p^{\gamma}}, \tag{12}$$

rather than the Maxwellian velocity distribution (which yields $exp(-x^2)$). In the case (12), the calculation yields the following asymptotic dependence of the Holstein function at large distances (here, the average velocity $v_T$ and lifetime $\tau$ refer to macroscopic perturbations of the medium):

$$T(R) \propto \frac{1}{R^{\gamma-5}}, \qquad \frac{R}{v_T\tau} \gg 1. \quad \gamma > 5, \tag{13}$$

To determine the relationship between pair correlations, one can easily derive the following scaling from the kinetic equation:

$$\frac{t}{\tau}\,T(R) \sim 1, \qquad \frac{R}{v_T\tau} \gg 1, \qquad \frac{t}{\tau} \gg 1. \tag{14}$$

It is evident that a value of γ − 5 = 2/3 corresponds to non-local transport of the Richardson law ($R^2 \propto t^3$).

It should be noted that the transport regime corresponding to resonant radiation transfer in the Lorentzian wing of the spectral line ($R \propto t^2$) is characterized by γ = 5.5, whereas transport in the Doppler wing ($R\sqrt{\ln(R)} \propto t$) corresponds roughly to γ = 6. It is also worth noting that while Richardson's law is reproduced in numerical calculations [12], there are regions where the observed scaling closely approximates the "Lorentzian" and "Doppler" cases.

## 5. Conclusions

Generalizing the results of [2] — which concern the modified Chapman-Enskog method and a nonlocal (i.e., non-diffusive, spatially integral) viscous stress tensor in the Navier-Stokes equation — to the case of arbitrary gradients of both hydrodynamic velocity and temperature has yielded a system of equations that simplifies the description of tangential discontinuities and separated flows. This simplification lies in the fact that the matching of the solution near discontinuities in the mean mass velocity can be performed without refining the computational grid, since the velocity gradient appears within an integral over spatial variables and is a relatively easily integrable function. Unlike the elementary matching of discontinuous solutions at a planar shock front, the description we have obtained allows for an accurate treatment of discontinuities (including non-tangential ones) for an arbitrary geometry of the discontinuity surface.

Applying the equation for the medium's mean mass velocity to the special case of an incompressible medium made it possible to derive a nonlocal vorticity transport equation that is structurally identical to the Biberman-Holstein equation for resonant radiation transport in plasmas and gases. This paved the way for solving the inverse problem: determining the form of the velocity distribution function for long-free-path vorticity carriers based on the known empirical Richardson $t^3$ law for the mean square of separation of a pair of particles in a turbulent liquid or gaseous medium. Measuring the characteristics of such medium perturbations is complicated by the difficulty of isolating them from the background of frequent perturbations with short free paths. The situation is analogous to the difficulty of measuring the far wings of a spectral line during resonant radiation transport. However, while the escape of photons from the medium enables the diagnosis of nonlocal energy transport, the diagnostics of nonlocality in the transport of hydrodynamic perturbations within a turbulent medium appear — in our view — to be less developed. We hope that the derived equation for nonlocal vorticity transport will prove useful for the numerical modeling and diagnosis of turbulence nonlocality.

The author is grateful to M.G. Levashova for her assistance with the numerical calculations of the Holstein function.